\documentclass[a4paper, traditabstract]{aa} 
\usepackage{graphicx}
\usepackage{txfonts}
\usepackage{natbib,twoopt}
\usepackage[breaklinks=true]{hyperref} 
\hypersetup{
  colorlinks   = true, 
  urlcolor     = red, 
  linkcolor    = blue, 
  citecolor    = blue, 
  breaklinks   = true 
}
\usepackage{bm}
\bibpunct{(}{)}{;}{a}{}{,}             

\newcommand*\diff{\mathop{}\!\mathrm{d}}
\newcommand*\Diff[1]{\mathop{}\!\mathrm{d^#1}}
\newcommand{\Planck}{{\em Planck}}

\begin{document}

   \title{A measurement of cluster masses using\\ \Planck\ and SPT-SZ CMB lensing}
   
   \author{Alexandre Huchet
          \and
          Jean-Baptiste Melin
          }

   \institute{Université Paris-Saclay, CEA, Département de Physique des Particules, 91191, Gif-sur-Yvette, France.\\
              \email{alexandre.huchet@cea.fr,jean-baptiste.melin@cea.fr}
                   }

   \date{Received ; accepted }

\abstract{We used an unbiased CMB lensing mass estimator on 468 SPT-SZ clusters from the SPT-SZ and the \Planck\ public data, the first such estimation using combined ground- and space-based data. We measured the average ratio between CMB lensing and SZ mass to be $M_{\rm CMBlens}/M_{\rm SZ} = 0.98 \pm 0.19 \rm{~(stat.)} \pm 0.03 \rm{~(syst.)}$. The average CMB lensing mass from the combination of the two data sets is measured at 4.8$\sigma$, which is a significant gain with respect to the measurement performed on the SPT-SZ only (3.9$\sigma$) or the \Planck\ only (3.7$\sigma$) data set. We showed that the combination not only takes advantage of the two different ranges of spatial scales (i.e. Fourier modes) observed but also exploits the lensing induced correlation between scales observed by one experiment and the other. This result demonstrates the importance of measuring a large range of spatial scales for CMB lensing mass estimation, from arcmin to degrees. This large range of scales will most probably be provided by the combination of various data sets, such as from the large and small aperture telescopes of the upcoming Simons Observatory and future CMB-S4 experiment, and \Planck. In this context, the \Planck\ data will remain a key element for CMB lensing cluster studies in the years to come.}

  \keywords{galaxies: clusters: general -- cosmic background radiation -- Gravitational lensing: weak -- methods: statistical}

\maketitle
%

\section{Introduction}

Clusters of galaxies, located at the nodes of the cosmic web, are very useful objects for studying both astrophysics and cosmology. They have played an important role in shaping and reinforcing the $\Lambda$CDM concordance model and have been used in many works to derive constraints on cosmological parameters~\citep[e.g.,][]{white_baryon_1993,carlberg_galaxy_1996,bahcall_constraining_1997,allen_cosmological_2002}. Among these approaches, the most promising way to constrain cosmological parameters with clusters is through cluster counts as a function of redshift and mass, which are very sensitive to cosmological parameters, particularly $\Omega_{\rm m}$ and $\sigma_8$~\citep{albrecht_report_2006,vikhlinin_chandra_2009,allen_cosmological_2011}. Clusters thus have the statistical power to be as efficient as other major cosmological probes but their constraints unfortunately suffer from uncertainty in the relation between cluster masses and directly observable quantities. This uncertainty on the cluster mass scale dominates current analyses~\citep[e.g.,][]{hasselfield_atacama_2013,planck_collaboration_xxiv_2016,bocquet_cluster_2019,abbott_dark_2020}. However, the situation will change in the next few years with the advent of the large optical facilities {\emph {Euclid}}\footnote{\url{https://www.esa.int/Science_Exploration/Space_Science/Euclid} \& \url{https://www.euclid-ec.org}} and the Vera Rubin Observatory\footnote{\url{https://rubinobservatory.org} \& \url{https://lsstdesc.org}}. These facilities will improve the accuracy and number of measurements of cluster masses thanks to weak gravitational lensing, up to redshifts of the order of one~\citep{laureijs_euclid_2011,abell_lsst_2009}. The major uncertainty for cosmology with clusters at $z<1$ will, therefore, be overcome in the next decade. Extending cosmological studies with clusters at redshift $z>1$ will not be possible with the weak gravitational lensing on galaxies due to the lack of background sources at these redshifts. The cosmic microwave background (CMB) is hoped to be a new background source for studying the weak lensing of clusters at higher redshift. Located at $z\sim1100$ and having precisely known statistical properties, it makes it possible to measure masses of galaxy clusters by weak gravitational lensing in the redshift range $0<z<3$. ~\citep{zaldarriaga_reconstructing_1999,seljak_lensing_2000,holder_gravitational_2004,dodelson_cmb_2004,vale_cluster_2004,lewis_cluster_2006,lewis_weak_2006}.

The first tools were proposed in the mid-2000s to detect this effect~\citep{maturi_gravitational_2005,hu_cluster_2007,yoo_improved_2008,yoo_lensing_2010}, but the first data sets for which the lensing signal became detectable only appeared in the early 2010s \citep{ruhl_south_2004,swetz_overview_2011,aghanim_planck_2020}. The signal-to-noise for CMB lensing mass measurement is very low in these data sets. Thus, it is not possible to make individual measurements and it is necessary to average the signal over several hundred clusters to pull the signal out of the noise. The first measurements were carried out almost jointly by the ACT, SPT and \Planck\ collaborations shortly before the mid-2010s~\citep{madhavacheril_evidence_2015,baxter_measurement_2015,planck_collaboration_xxiv_2016}. These detections were made possible thanks to the pioneering work cited above and improved tools~\citep[see][]{melin_measuring_2015}. The improvement of existing tools and the development of new tools are currently active research fields~\citep{raghunathan_measuring_2017,madhavacheril_mitigating_2018,raghunathan_inpainting_2019,horowitz_reconstructing_2019,patil_suppressing_2020,gupta_mass_2021,levy_foreground_2023,chan_small_2023}. The motivation is to be ready to analyze data from the new generation of CMB instruments which will provide data sets that will enable individual mass measurements for the first time, the Simons Observatory~\citep[SO,][]{ade_simons_2019} and CMB-S4~\citep{abazajian_cmbs4_2019}.
Following its first detection in the mid-2010s, the weak gravitational lensing on the CMB has been used to measure masses of clusters selected in the optical and the infrared~\citep{geach_cluster_2017,baxter_measurement_2018,raghunathan_mass_2019,madhavacheril_atacama_2020} as well as to measure the mass of halos hosted by galaxies or quasars~\citep{raghunathan_imprints_2018,geach_halo_2019}. These measurements are performed using the intensity (i.e. temperature) maps of current data sets, but the polarized signal is expected to improve significantly the precision of the measurement in future data sets. In particular, \cite{raghunathan_detection_2019} recently succeeded in making the first detection of the CMB cluster lensing in the SPTpol data using only the polarized signal. In parallel to these efforts, first cluster cosmological analyses using CMB cluster lensing to anchor the mass scale were performed~\citep{planck_collaboration_xxiv_2016,zubeldia_cosmological_2019}, including a careful assessment of the statistics of the recovered CMB lensing masses~\citep{zubeldia_quantifying_2020}. With the upcoming experiments (SO and CMB-S4), it will be possible to extract, from the same data set, galaxy clusters and their masses to high accuracy thanks to CMB cluster lensing, leading to competitive cosmological constraints~\citep{louis_calibrating_2017,madhavacheril_fundamental_2017,raghunathan_constraining_2022,chaubal_improving_2022}.

In this work, we used an unbiased CMB lensing mass estimator on 468 clusters from the SPT-SZ cluster catalogue \citep{bocquet_cluster_2019}. We worked on flat tangential maps cut from the all-sky \Planck\ and SPT-SZ maps, and combined them. We took advantage of the different ranges of spatial scales, or Fourier modes of the maps, that the two data sets provide. The combined signal-to-noise ratio for individual mass measurement is comprised between 0.05 and 1. We thus had to average the 468 mass estimations to obtain the average mass of the whole sample.

We first introduce the two data sets used in our analysis in
Sect.~\ref{sec:data}. We then explain our methodology in Sect.~\ref{sec:methodo} and present our simulations in~Sect. \ref{sec:simu}. We show our results in~Sect.~\ref{sec:res_data}. We discuss uncertainties in Sect.~\ref{sec:uncer}. Finally, we summarize and conclude in Sect.~\ref{sec:conclu}.

Throughout the article, we work in the flat $\Lambda$CDM model with parameters
$H_0 = 70$~km/s/Mpc, $\Omega_{\rm m} = 0.3$.

\section{Data sets} \label{sec:data}

    We used two sets of sky maps in our analysis, SPT-SZ and \Planck.
    Our analysis is performed on the SPT-SZ cluster catalogue only.

\subsection{SPT-SZ sky maps}

    The South Pole Telescope \citep[SPT,][]{ruhl_south_2004} is a 10~m diameter mirror facility operating at the Amundsen–Scott South Pole Station in Antartica. It observed
    a $\sim$2500~deg$^2$ area of the southern sky between 2008 and 2011,
    in three frequency bands centered around 95, 150 and 220~GHz, with a
    beam size of 1.6, 1.1 and 1.0~arcmin respectively (full width at
    half maximum, FWHM).This survey is referred as the SPT-SZ survey \citep{story_measurement_2013}. SZ stands for
    the Sunyaev-Zel'dovich effect, used to detect the clusters
    in this survey. The SZ effect is actually a contaminant in our study, we
    present it and explain how we handle it in Sec.~\ref{subsec:ILC}.

    We used the "SPT Only Data Maps" published by~\cite{chown_maps_2018}. The maps include all three SPT-SZ frequencies,
    with a degraded resolution compared to the original data, that is
    1.75~arcmin (FWHM) for the three maps. The SPT-SZ data contain a point source mask and the transfer
    function of the instrument for the three frequencies. All maps are provided in equatorial coordinates, and 
    in HEALPix format with the resolution parameter $N_{side} = 8192$, corresponding to a pixel
    size of about 0.43 arcmin. We also extracted the frequency response of each of the SPT-SZ band from Fig. 10 of~\cite{chown_maps_2018}, which are useful to compute the frequency dependance of the thermal SZ effect in each frequency band $j_{\nu_i}$ (see Sect.~\ref{sec:overview}).
   
\subsection{Planck sky maps}

    The \Planck\ space mission~\citep{aghanim_overview_2020} is equipped with a 1.5~m diameter telescope and two instruments,
    the Low Frequency Instrument (LFI) and the High Frequency Instrument (HFI). The
    latter observed the whole sky in six frequency bands centered at 100, 143,
    217, 353, 545 and 857~GHz at a resolution of 9.659, 7.220, 4.900, 4.916,
    4.675, 4.216~arcmin (FWHM) \citep{ade_second_2016}.
  
     \Planck\ HFI maps are full sky but we restrained our
    analysis to the SPT-SZ field in order to combine the maps.
    
    The Planck maps are provided in Galactic coordinates, at a HEALPix resolution $N_{side} = 2048$, corresponding to a pixel size of 1.72~arcmin. To facilitate combining these maps with the SPT-SZ maps, we changed their coordinate system to equatorial, and upgraded them to $N_{side} = 8192$ by zero-padding: we harmonic transform the maps, add zeros at high multipoles in the harmonic space, and then perform the inverse harmonic transform. We used the maps from the \Planck\ Release 2 provided in the \Planck\ Legacy Archive.

\subsection{Complementarity of the two data sets}

    The two data sets are complementary.
    On the one hand, the \Planck\ telescope is located in space, and its HFI
    mapped the sky at six different frequencies, but \Planck\ has somewhat a small mirror for cluster science.
    On the other hand, the South Pole Telescope is ground-based, has a large mirror, a large number of detectors, but the large scales in the maps  ($> 0.5 {\rm deg}$) are filtered out 
    due to the telescope scanning strategy.
    
    As a result, SPT-SZ has a very good resolution and sensitivity whereas
    \Planck\ has an excellent frequency coverage and can map
    the large angular scales of the sky. This complementarity should bring information
    on correlations between large (\Planck) and small (SPT-SZ) scales,
    leading to a significant improvement on the final error of our combined CMB lensing analysis compared to the analyses led on the two independent data sets.

\subsection{SPT-SZ cluster catalogue}
\label{sec:sptszcat}

\begin{figure}
    \centering
    \includegraphics[width=\hsize]{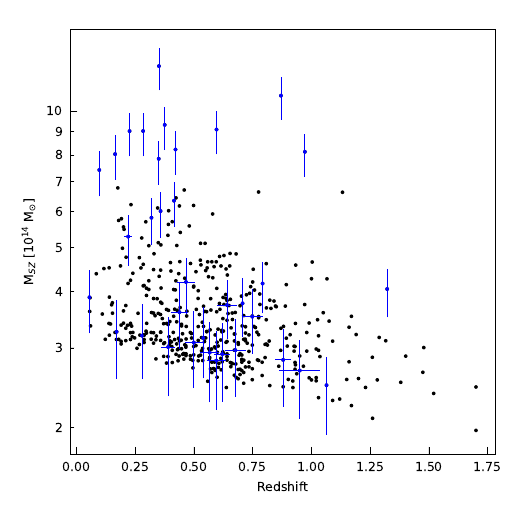}
      \caption{Mass-redshift distribution of the 468 SPT-SZ clusters used in our analysis. We display in blue the error bars for the clusters with $M_{\rm SZ} > 7 \times 10^{14} M_\sun$ and of one out of 20 under this threshold for clarity.}
         \label{fig:SPT_SZ_clu}
    \end{figure}

    We used the catalogue from \citet{bleem_galaxy_2015}, with masses and redshifts updated by~\citet{bocquet_cluster_2019}. The catalogue contains 677 objects. We restricted our analysis to the 516 clusters with measured redshifts, and subsequently removed 48 clusters located close to the edge of the SPT-SZ field (see Sect.~\ref{subsec:tan_maps}). Our final catalogue contains 468 clusters. The SPT-SZ masses $M_{500}$ (mass enclosed in a sphere of radius $R_{500}$ within which the average density is 500 times the critical density at the cluster's redshift) are estimated using the SZ signal-to-noise ratio versus mass relation for a flat $\Lambda {\rm CDM}$ model ($h=H_0/(100 \, {\rm km/s/Mpc})=0.7$, $\Omega_{\rm m}$=0.3, $\sigma_8=0.8$), which is calibrated from external X-ray and weak lensing data sets. The SPT-SZ masses are corrected from selection effects. We multiplied the SPT mass by a factor 0.8 to match the \cite{arnaud_universal_2010} mass definition as was done by~\cite{tarrio_comprass_2019}, \cite{melin_pszspt_2021} and \cite{melin_measurement_2023}. We denote by  $M_{\rm SZ}$ the SPT mass rescaled by this factor 0.8. We discuss this choice in the light of our results in Sect.~\ref{sec:conclu}. The distribution of our sample in the mass versus redshift plane is shown in Fig.~\ref{fig:SPT_SZ_clu}. Clusters span a large redshift range $z \in [0,1.7]$ above a mass threshold that varies slowly with redshift $M_{\rm SZ} \gtrsim 2 \times 10^{14} M_\sun$.

\section{Methodology} \label{sec:methodo}

    We use some of the notations from \citet{melin_measuring_2015}
    in the following subsections.
    
\subsection{Concise overview}
\label{sec:overview}

    Our goal is to use an unbiased estimator to measure the mean $M_{\rm CMBlens}/M_{\rm SZ}$ mass ratio. With this in mind, we did not stack the
    maps of all the clusters but instead built
    a CMB map for each cluster field, from the nine frequency
    maps (three SPT and six \Planck).
    
    We first cut tangential maps out of
    the nine sky maps centered at the location of each cluster.
    We then used an Internal Linear Combination (ILC,
    Sect.~\ref{subsec:ILC}) to build the best lensed CMB
    map around the cluster. Gravitational lensing of the CMB induces correlations across its spatial scales. We used these correlations
    to build a lensing potential with the minimum variance
    estimator from \cite{hu_mass_2002}. This lensing potential profile
    was then compared to the lensing potential of a Navarro-Frenk-White~\citep[NFW,][]{navarro_structure_1996} profile using a matched
    filter \citep{melin_measuring_2015}. We can then derive the individual mass from the NFW profile.
    Individual errors are large (typically one to ten times larger than the signal) so
    we computed a weighted mean mass of the sample as our measurement.

    All those steps were applied on eleven different sets of
    468 locations. 
    The first set is the one with the
    CMB cluster lensing signal. We called it the \emph{on} set. It is centered at the locations of the clusters. The
    ten following sets are ten draws at 468 random locations in
    the SPT-SZ footprint. We called them the \emph{off} sets. We ran the same analysis on the \emph{on} and
    \emph{off} sets of maps, and subtracted at the end the average signal of the ten \emph{off} sets from the \emph{on} set:  $\emph{on}- \langle \emph{off} \rangle_{10}$.
    This method allowed us to remove spurious signal from instrumental correlated noises, foreground or background astrophysical sources and any additive systematic bias.
    Averaging ten random sets reduces the impact on the final error, which takes the form  $\sigma = \sqrt{\sigma_\emph{on}^2 + \sigma_\emph{off}^2} = \sqrt{\sigma_\emph{on}^2 (1+\frac{1}{10})}$.
    
    We assumed that each field contains the primary CMB (lensed!), the thermal SZ (tSZ) signal, and noise. The noise can be of astrophysical or instrumental origin. We thus write the maps of one field in Fourier space as 

    \begin{equation}
      \label{eq:master_indiv_eq}
      \left\{\begin{array}{@{}l@{}}
        m_{\nu_0}(\bm{k}) = \alpha_{\nu_0}(\bm{k})S(\bm{k}) + \beta_{\nu_0}(\bm{k})y(\bm{k}) + n_{\nu_0}(\bm{k})\\
        m_{\nu_1}(\bm{k}) = \alpha_{\nu_1}(\bm{k})S(\bm{k}) + \beta_{\nu_1}(\bm{k})y(\bm{k}) + n_{\nu_1}(\bm{k})\\
        ...
      \end{array}\right.,
    \end{equation}

   or in a more compact way
   
     \begin{equation}
     \label{eq:master_eq}
        \bm{m} (\bm{k}) = \bm{\alpha}(\bm{k})S(\bm{k}) + \bm{\beta}(\bm{k})y(\bm{k}) + \bm{n}(\bm{k}).
    \end{equation}
   
    with $S$ the primary CMB, $y$ the tSZ Compton map of the cluster, and $n_{\nu_i}$ the instrumental and astrophysical noises, that is
    all the other components in the frequency band $\nu_i$. We also have

    \begin{equation}
    \label{eq:alpha_beta}
      \left\{\begin{array}{@{}l@{}}
        \alpha_{\nu_i}(\bm{k}) = b_{\nu_i}(k) \times t_{\nu_i}(\bm{k}) \\
        \beta_{\nu_i}(\bm{k}) = \alpha_{\nu_i}(\bm{k}) \times j_{\nu_i}
      \end{array}\right.\,,
    \end{equation}
    
    \noindent
    where $b_{\nu_i}$ and $t_{\nu_i}$ are respectively the isotropic beam and
    the transfer function of the instrument in the frequency band $\nu_i$. This
    transfer function is taken as $t_{\nu_i} = 1$ for all \Planck\ frequencies
    but is anisotropic for SPT-SZ frequencies. Therefore, $\alpha_{\nu_i}$
    and $\beta_{\nu_i}$ depend on the vector $\bm{k}$ and not only on its modulus
    for SPT-SZ data, making the analysis more complex. $j_{\nu_i}$ is the
    frequency dependence of the tSZ effect, integrated in the $\nu_i$ band.

    The kinetic SZ (kSZ) signal has the same frequency dependence as the primary CMB and cannot be separated from it. We estimated its impact on our result in Sect.~\ref{sec:kSZ} and corrected from it in our final measurement.

\subsection{Tangential maps}
\label{subsec:tan_maps}

   We performed our analysis in Fourier space from small flat tangential maps. We hence had to cut them out of the spherical HEALPix maps.
   
   For each cluster, we cut a 10x10~deg$^2$ map centered on it, when possible.
   This size matches the size used for the cluster extraction in the \Planck\ 
   official analysis~\citep{ade_second_2016}. It is large enough to allow for a good estimation of the covariance matrix of the maps. If the cluster was
   close to the border of the field of the SPT-SZ public data, that is less than 17 arcmin
   ($\sim$40 pixels), we shifted the cluster as close
   as possible to the center without including in the tangential maps more than 1\% of bad pixels, that is empty pixels out of the SPT-SZ field. Some clusters being too close to the border,
   we had to remove them from our sample. We removed 48 clusters, reducing the SPT-SZ sample from 516 clusters with redshift to 468 clusters with redshift and well within the SPT-SZ public field. The SPT transfer function is not azimuthally symmetric. We estimated it as in~\cite{melin_pszspt_2021}. We computed it, for each frequency, at the location of the 468 SPT clusters under the form of an HEALPix map. We then cut 10x10~deg$^2$ maps around the cluster location, averaged them and symmetrised over the horizontal and vertical directions to obtain a single transfer function per frequency. Throughout the analysis, we apodised the maps in real space to avoid artifacts when using the (2-dimensional) Fourier transform.

\subsection{Point sources handling}

    The tangential maps contained point sources that we needed to remove to avoid contamination of our lensing estimator.
    To do so, we replaced a circular zone
    slightly larger than the FWHM of the beam centered at the point source location
    by a constrained Gaussian realisation. This filling had to be
    continuous with the surrounding data in order to reduce the
    impact on our lensing estimator.

    For each cluster, we had 9 maps -- 6 \Planck\ and 3 SPT-SZ
    -- that we masked separately. We first computed the signal-to-noise ratio of the point sources in
    each map using a single frequency matched filter. We then iterated on the
    sources verifying $S/N>6$ and masked them with a disk of radius $r_i$ of 3 (resp. 10)~arcmin for SPT-SZ (resp. \Planck).

    To mask the point sources, we could not simply set the masked regions to zero, no
    matter how smooth their edges were. We later measured correlations
    between different k-modes (spatial scales) of the map, and these masked regions
    would only have created fake correlations if filled with zeros. To ensure that we were not creating
    artificial correlations, we used the \citet{hoffman_constrained_1991}
    constrained Gaussian field method \citep[see also][]{benoit-levy_full-sky_2013}.
      
\begin{figure}
    \centering
    \includegraphics[width=\hsize]{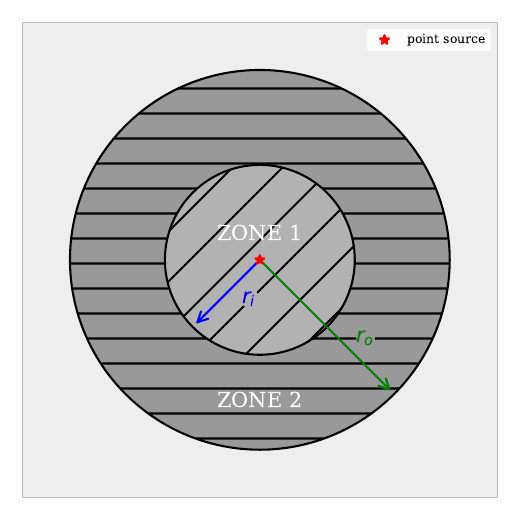}
      \caption{Diagram of the two zones used for in-painting. Zone 1 is the in-painted area
      (the masked zone), and zone 2 is
      the area constraining our Gaussian field to ensure continuity
      (the constraining zone). The inner radius $r_i$ is 3 (resp. 10)
      arcmin for SPT (resp. \Planck) and the outer
      radius $r_o$ is twice the inner one.
              }
         \label{fig:schema_fill}
    \end{figure}

    We first computed a random realisation of a Gaussian temperature
    field $\tilde{m}$ from the 2-dimensional power spectrum $P(\bm{k})$
    of the map to be masked $m$. This random realisation was expected to fill the
    hole and, although it already had the same variance as $m$,
    it needed to be continuous with the neighbouring original
    map pixels. We thus had to constrain it, using an annulus encircling
    the masked zone named the calibrating zone. We chose it with an
    outer radius $r_o$ of twice the inner radius $r_i$, the radius of the masked
    zone. We illustrate the disk and annulus in Fig.~\ref{fig:schema_fill}.

    We now had a zone to mask with a field $m_1$ and a constraining zone
    with a field $m_2$ in the original map, and the same for the Gaussian
    realisation ($\tilde{m}_1$ and $\tilde{m}_2$).

    The second step involved computing the 2-dimensional correlation function
    of $m$, that we noted $\xi(\bm{r})$ with $\bm{r}$ the distance
    between two pixels. $\xi(\bm{r})$ is the inverse Fourier transform of $P(\bm{k})$, and describes the most likely value of a
    pixel given the values of other pixels at different distances. From this
    function, we can obtain the covariance matrix for chosen sets
    of pixels. We created two of them, one being the auto-covariance matrix of the
    constraining zone $\Sigma_{22}$ and the other one the covariance matrix between
    the masked zone and the constraining one $\Sigma_{12}$. We understand now why we
    used the correlation function on the whole map to compute those, as number of pixels in the constraining zone might not be enough to compute $\Sigma_{22}$ accurately and, since the values of $m_1$
    are the ones we have to get rid of, we cannot use only them to compute
    $\Sigma_{12}$.

    The conditional probability distribution function of a field $m'_1$
    constrained by $m'_2$ is

    \begin{equation}
        \mathcal{P}(m'_1|m'_2) = \frac{\mathcal{P}(m'_2|m'_1)\mathcal{P}(m'_1)}{\mathcal{P}(m'_2)},
    \end{equation}
    
    \noindent
    and it is a Gaussian centered on

    \begin{equation}
        \bar{m}'_1 = \langle \mathcal{P}(m'_1|m'_2) \rangle = \Sigma_{12}\Sigma_{22}^{-1}m'_2.
    \end{equation}

    $\bar{m}'_1$ can be seen as the most probable value for each pixel in
    zone 1 \emph{individually}, i.e. what we should get if we know $m'_2$ and $\xi(\mathbf{r})$. This does not mean that $\bar{m}'_1$ had the same
    statistical properties as the CMB we wanted to reconstruct: it lacked the
    deviations between the values in zone 1 inferred from zone 2 only and the actual values. These deviations that describe the large
    scales that we cannot constrain from zone 2 were obtained from the simulation.
    
    Our field $\tilde{m}_1$ already contained the appropriate random variations
    because we built it with $P(\bm{k})$, but it needed to be offset
    to have the same mean as $m_1$ should have.

    The third step was to compute the residual of the realisation,
    defined as

    \begin{equation}
        m_{1,r} = \tilde{m}_1 - \bar{\tilde{m}}_1 = \tilde{m}_1 - \Sigma_{12}\Sigma_{22}^{-1}\tilde{m}_2,
    \end{equation}

    \noindent
    which is the field with only the random variations from the most
    probable zone 1, $\bar{\tilde{m}}_1$ (induced from the correlations with zone 2).

    The last step was then to add $\bar{m}_1$, that is the mean $m_1$
    should have:

    \begin{equation}
        m_{1,fill} = m_{1,r} + \bar{m}_1 = \tilde{m}_1 + \Sigma_{12} \Sigma_{22}^{-1} (m_2 - \tilde{m}_2).
    \end{equation}

    $m_{1,fill}$ is used to fill the zone to mask, it is continuous
    with the surrounding pixels and has the expected standard deviation.

\subsection{Internal linear combination}
\label{subsec:ILC}

    We had nine point-source-free maps for each cluster,
    but they were still contaminated by the tSZ effect~\citep{carlstrom_cosmology_2002}.
    The maps were written in column in the from of~Eq.~\ref{eq:master_eq}.

    We clearly see where our signal lies and we look
    for a minimum variance estimate $\hat{S}$ of the pure lensed CMB map $S$.
    We used the method provided by \citet{remazeilles_cmb_2011}, called
    constrained Internal Linear Combinations (ILC) to obtain our minimum variance estimate while nulling the tSZ signal. As its name
    suggests, its aim is to recover the signal thanks to
    a linear combination of the frequency maps, i.e.

    \begin{equation}
        \hat{S} = \bm{w}^t \bm{m}.
    \end{equation}

    A derivation of the constraints $\bm{w}^t \bm{\alpha} = 1$
    and $\bm{w}^t \bm{\beta} = 0$ gives

    \begin{equation}
        \bm{w}^t = \frac{(\bm{\beta}^t \hat{\bm{R}}^{-1} \bm{\beta}) \bm{\alpha}^t \hat{\bm{R}}^{-1} - (\bm{\alpha}^t \hat{\bm{R}}^{-1} \bm{\beta}) \bm{\beta}^t \hat{\bm{R}}^{-1}}
        {(\bm{\alpha}^t \hat{\bm{R}}^{-1} \bm{\alpha}) (\bm{\beta}^t \hat{\bm{R}}^{-1} \bm{\beta}) - (\bm{\alpha}^t \hat{\bm{R}}^{-1} \bm{\beta})^2},
    \end{equation}
    
    \noindent
    where $\hat{\bm{R}}$ is the empirical covariance matrix of the nine
    maps. In their work, \citet{remazeilles_cmb_2011} compute
    $\hat{\bm{R}}$ with an average on $l$-intervals in spherical harmonics
    in order to limit the statistical variations. In our case, because the SPT transfer function is strongly anisotropic,
    we could not use the same method. We computed $\hat{\bm{R}}$
    on all the \emph{on} and \emph{off} fields
    and performed the average on individual $\bm{k}$ modes to preserve the
    maximum of information. We  averaged indiscriminately the \emph{on}
    and \emph{off} maps because the CMB lensing signal is not expected to affect
    the ILC reconstruction of the CMB when averaged over all the maps.

     This method allowed us to recover the best CMB from the nine frequency maps,
    but the noise at small scales was very large in the final map due to the deconvolution of the beams and transfer functions $\bm{\alpha}$ and $\bm{\beta}$. We thus applied an effective beam $\tilde{b}$ to the map that flattened the
    noise at small scales. This effective beam was taken into account in the lensing estimator
    in Sect.~\ref{subsec:lens_pot_est}. For the \Planck\ only analysis, we used a Gaussian beam with a FWHM of 4.9~arcmin, whereas for the SPT-SZ only analysis we used the value of $\alpha_{\nu_i}(k)$ that had the smallest modulus mode-wise out of the three frequencies. For the joint analysis, the effective beam was taken as the largest modulus mode-wise between the \Planck\ effective beam and the SPT-SZ effective beam in Fourier space.

    The mean power spectrum of the final map convolved by the effective beam is:

    \begin{equation} \label{eq:ILC_var}
        P_{\hat{S}} = \langle |\tilde{b} \bm{w}^t \bm{m} |^2 \rangle = |\tilde{b}|^2
        \frac{(\bm{\beta}^t \hat{\bm{R}}^{-1} \bm{\beta})^2 (\bm{\alpha}^t \hat{\bm{R}}^{-1} \bm{\alpha})
        - (\bm{\alpha}^t \hat{\bm{R}}^{-1} \bm{\beta})^2 (\bm{\beta}^t \hat{\bm{R}}^{-1} \bm{\beta})}
        { \left [ (\bm{\alpha}^t \hat{\bm{R}}^{-1} \bm{\alpha}) (\bm{\beta}^t \hat{\bm{R}}^{-1} \bm{\beta}) - (\bm{\alpha}^t \hat{\bm{R}}^{-1} \bm{\beta})^2 \right ]^2},
    \end{equation}

    \noindent
    where $\langle \rangle$ is the ensemble average over various hypothetical sky realisations. $P_{\hat{S}}$ is used in the definition of the lensing potential operator in the next section.
      
    This method was used on the maps of the nine frequencies for
    $|\bm{k}|<175$ (or equivalently $l<6287$), i.e. when still in the original \Planck\
    resolution. For $175 \le |\bm{k}| < 278$, where
    \Planck\ does not provide information, we only used the
    three maps from SPT. At $|\bm{k}| \ge 278$, equivalent
    to $l>10000$, neither SPT nor \Planck\ provide any reliable
    information, we thus masked the corresponding ILC $\bm{k}$ modes.
   
\subsection{Lensing potential estimation} \label{subsec:lens_pot_est}

    We then had to recover the lensing signal from the pure lensed CMB map obtained in the previous section. This signal is linked to the
    gravitational potential, directly related to the mass projected along the line of sight, and can
    therefore be used to measure the mass of the cluster.
    We chose the \citet{hu_mass_2002} minimum variance lensing
    estimator to do so. This estimator uses the fact that
    the CMB lensing induces correlations between the CMB $\bm{k}$ modes.
    They define the estimator in spherical harmonics but we used
    the flat-sky equivalent, as done in \cite{melin_measuring_2015}.
    The deflection (or lensing) potential $\hat{\phi}$ was computed in Fourier
    space as follows:

    \begin{equation}
        \hat{\phi}(\bm{K}) = A(\bm{K}) \sum_{\bm{k}} \hat{S}^*(\bm{k}) \hat{S}(\bm{k'}) F(\bm{k},\bm{k'}),
    \end{equation}

    \noindent
    where $\bm{K} \equiv \bm{k} - \bm{k'} \pmod{n}$, with $n$
    the size of the image in pixels along the x or y axis.
    The normalisation $A$, that is also the variance of the
    reconstructed lensing potential, is defined as

    \begin{equation}
        A(\bm{K}) = \left[ \sum_{\bm{k}} f(\bm{k},\bm{k'}) F(\bm{k},\bm{k'}) \right]^{-1},
    \end{equation}
    
    \noindent
    with the weights $F$ defined to minimise the variance of the
    estimator $\hat{\phi}$:

    \begin{equation}
        F(\bm{k},\bm{k'}) = \frac{f^*(\bm{k},\bm{k'})}{2 P_{\hat{S}}(\bm{k}) P_{\hat{S}}(\bm{k'})},
    \end{equation}

    \noindent
    $P_{\hat{S}}$ being the mean of the ILC power spectrum defined
    in Eq.~\ref{eq:ILC_var}. The minimum variance filter $f$ is defined
    as:

    \begin{equation}
        f(\bm{k}, \bm{k'}) = \tilde{b}^*(\bm{k}) \tilde{b}(\bm{k'})
        \left[ \tilde{C}_k \bm{k} . \bm{K} - \tilde{C}_{k'} \bm{k'} . \bm{K} \right],
    \end{equation}

    \noindent
    with $\tilde{b}$ the effective instrumental beam applied to the combination
    of the nine maps (Sect.~\ref{subsec:ILC}). $\tilde{C}_k$ is the power spectrum of the unlensed CMB. We obtained it
    from an all-sky pure CMB Gaussian simulation at HEALPix resolution $N_{side} = 8192$, drawn from the \Planck\
    primary CMB power spectrum, that we projected and apodised as we did for the sky maps (Sect.~\ref{subsec:tan_maps}).

\subsection{Matched filter}
\label{subsec:mfilt}

    We then wanted to measure a mass from the reconstructed lensing profile. To this end, we used the matched filter
    described in \citet{melin_measuring_2015}:

    \begin{equation} \label{eq:mfilt}
        M_{\rm CMBlens}/M_{\rm fiducial} = \left[ \sum_{\bm{K}} \frac{|\Phi(\bm{K})|^2}{A(\bm{K})} \right]^{-1}
        \sum_{\bm{K}} \frac{\Phi^*(\bm{K})}{A(\bm{K})} \hat{\phi}(\bm{K}),
    \end{equation}

    \noindent
    where $\Phi$ is the lensing potential of a Navarro-Frenk-White (NFW)
    density profile \citep[see][]{navarro_structure_1996} for a given
    fiducial cluster mass $M_{\rm fiducial}$. We detailed this profile and its use in Sect.~\ref{subsec:clu_defl_pot}.

    The matched filter returns the amplitude of our mass measurement
    with respect to the chosen fiducial mass and can, because of
    a low S/N, return a negative amplitude and as a consequence a negative mass. Because the individual error bars were large, we averaged the individual measurement over the full sample of 468 clusters.
    Performing individual measurements also allowed us to check for redshift and mass dependence of our $M_{\rm CMBlens}/M_{\rm fiducial}$.

\section{Simulations} \label{sec:simu}

    We first tested the method on simulated data to study the statistics of the recovered mass and assess for possible biases. Therefore, we built nine
    lensed CMB maps for each cluster, with the instrumental characteristics of the nine frequencies.

\subsection{Cluster deflection potential} \label{subsec:clu_defl_pot}

    We chose a NFW density profile (\citet{navarro_structure_1996}) for the cluster,

    \begin{equation}
        \rho (r) = \frac{\delta_c \rho_{\rm c}}{(r/r_s) (1 + r/r_s)^2},
    \end{equation}
 
    \noindent
    where $\rho_{\rm c}$ is the critical density at redshift $z$ and $r_s = R_{500}/c_{500}$
    the scale radius, with $R_{500}$ the radius inside which the cluster
    density is 500 times $\rho_{\rm c}$. $c_{500}$ is
    the concentration parameter that we assumed to be constant
    $c_{500}=3$ for simplicity, although it is expected to weakly vary with the
    mass and redshift of the cluster \citep[e.g.][]{diemer_universal_2015,ludlow_mass_2016}. The characteristic
    overdensity $\delta_{\rm c}$ is linked to $c_{500}$ as
    follows:

    \begin{equation}
        \delta_{\rm c} = \frac{500}{3} \frac{c_{500}^3}{\ln(1+c_{500})-c_{500}/(1+c_{500})}.
    \end{equation}

    We projected this density profile along the line of sight up to a
    distance of $5 R_{500}$ from the center of the cluster and obtained
    a surface density profile

    \begin{equation}
        \Sigma_{NFW}(r_{\perp}) = \int_{los} \rho(r_{\perp},l) \diff l.
    \end{equation}

    \noindent
    
    We followed \citet{bartelmann_weak_2001} to obtain the lensing profile from the surface mass density of the
    cluster. We first got the convergence

    \begin{equation}
        \kappa (\bm{\theta}) = \frac{\Sigma_{NFW}(D_d \bm{\theta})}{\Sigma_{\rm crit}},
        \; \mathrm{where} \; \Sigma_{\rm crit} = \frac{c^2}{4 \pi G} \frac{D_s}{D_d D_{ds}},
    \end{equation}

    \noindent
    with $\bm{\theta}$ the angular distance to the center of the cluster. $\Sigma_{\rm crit}$
    is the critical surface mass density above which the lensing produces
    multiple images, i.e. it separates weak and strong lensing. $D_d$, $D_s$ and $D_{ds}$,
    are respectively the distances observer-lens (i.e. observer-cluster), observer-source (i.e. observer-CMB) and
    lens-source (i.e. cluster-CMB).

    From the convergence $\kappa$, we derived the
    lensing potential

    \begin{equation}
        \Phi (\bm{\theta}) = \frac{1}{\pi} \int_{\mathbb{R}^2} 
        \kappa(\bm{\theta'}) \ln |\bm{\theta} - \bm{\theta'}| \Diff{2}{\theta'}.
    \end{equation}

    This potential was used in the simulation of the
    lensed CMB maps, but also in the matched filter from Eq. \ref{eq:mfilt}.
    It does not depend on the observed frequency. We used it for
    building the lensed maps.

\subsection{Thermal SZ distortion}
\label{subsec:SZ_effect}
    
    We also needed to model the thermal SZ signal from the cluster. This signal
    is proportional to the electron pressure of the intracluster
    gas \citep{carlstrom_cosmology_2002}. We used the approach
    of \citet{nagai_effects_2007} and \cite{arnaud_universal_2010}, with a generalized NFW
    electron pressure profile (gNFW):

    \begin{equation}
    \label{eq:press_prof}
        P(r) = \frac{P_{500}P_0}{(r/r_s)^{\gamma}[1 + (r/r_s)^{\alpha}]^{(\beta - \gamma)/\alpha}},
    \end{equation}

    \noindent
    where ($\gamma$, $\alpha$, $\beta$) are the slope of the profile
    at distances $r \ll r_s$, $r \sim r_s$ and $r \gg r_s$ respectively.
    $P_0$ is the normalisation and $P_{500} \propto M_{500}^{5/3}$ is
    a characteristic pressure with $M_{500}$ the cluster mass inside $R_{500}$.
    The SZ signal is used to detect clusters and to estimate masses in combination with external data~\citep{bocquet_cluster_2019}.
    We used these $M_{\rm SZ}$ masses (rescaled with a factor 0.8 as discussed in Sect.~\ref{sec:sptszcat}) as the fiducial mass $M_{\rm fiducial}$ to simulate our clusters (dark matter NFW, tSZ) and build the lensing profile for the matched filter.
        
    We used the \citet{arnaud_universal_2010} values for the parameters in Eq.~\ref{eq:press_prof}
    and derived the Compton parameter $y$:

    \begin{equation}
        y(\theta) = \frac{\sigma_T}{m_e c^2} \int_{los} P(D_d\theta,l) \diff l,
    \end{equation}

    \noindent
    with $\sigma_T$ the Thomson cross-section and $m_e c^2$
    the electron rest mass energy.
    
    The SZ temperature distortion $\Delta T_{SZE}$ at a frequency
    $\nu$ and a distance $r$ from the center of the cluster is then
    given by
    
    \begin{equation}
        \frac{\Delta T_{SZE}}{T_{CMB}}(\nu,\theta) = j_{\nu} y(\theta),
    \end{equation}
   
    \noindent
    where we see the separated dependence of the tSZ effect in position
    $\theta$ and frequency $\nu$. We included relativistic corrections to $j_{\nu}$ following Sect. 3.3 of~\citet{melin_measurement_2023}. We used the cluster mass-temperature from Eq. 19 of~\citet{bulbul_xray_2019} for each cluster and the formula of~\citet{itoh_relativistic_1998}. The tSZ spectrum including the relativistic correction was integrated over the spectral bandwidth of each of the nine frequency bands.

\subsection{Final maps}

    For each frequency map, we put the same realisation of CMB $\tilde{S}$,
    generated from the CMB power spectrum provided by \citet{planck_collaboration_xiii_2016}.
    We then used the cluster deflection potential
    to obtain the lensed CMB $S$~\citep{bartelmann_weak_2001}.

    \begin{equation}
        S = \tilde{S} + \nabla \tilde{S} . \nabla \Phi .
    \end{equation}

    This lensed CMB, common between all frequencies was then altered
    with frequency or instrumental dependant elements. We added the
    SZ effect $\Delta T_{SZE}$ in the maps and convolved them
    with the corresponding beam, before adding the appropriate
    noise, the same way we defined the maps in Eq.~\ref{eq:master_indiv_eq}. For the simulations, we assumed that the instrumental noise is white with levels of 40, 18 and 70 $\mu$K.arcmin at 95, 150 and 220~GHz for SPT-SZ, and 38, 31, 46, 1.4$\times 10^2$, 1.3$\times 10^3$, 5.5$\times 10^4$ $\mu$K.arcmin at 100, 143, 217, 353, 545 and 857~GHz for \Planck.

\subsection{Results on simulations}

    For each of our 468 SPT-SZ clusters, we simulated
    nine frequency maps with the cluster CMB lensing signal ($\Phi$
    and $\Delta T_{CMB}$), the \emph{on} maps. We also simulated ten
    sets of nine frequency maps without the CMB lensing signal as the \emph{off}
    maps. By doing \emph{on} - $\langle \emph{off} \rangle_{10}$, we removed 
    spurious signal from instrumental correlated noises, foreground or background astrophysical sources and any additive systematic bias. In the simulations, we first created periodic maps, that is
    where the edges have continuity with the opposite one. We did not apodise the periodic maps. We also placed clusters at the center of the maps to avoid any systematic effect linked to
    non periodicity. In this scenario, we found no bias in our measurement, and
    we did not need to remove the average of the \emph{off} maps to estimate the masses.
    It shows that the method itself does not suffer from any bias when the true lensing profile is known and when the maps do not contain any 
    correlated noise, or foreground/background sources.

    This case, however, is not fully realistic. Some of our clusters
    were not centered in the maps and the CMB is certainly not continuous from
    one edge to the opposite one. If a cluster is not centered, it has a non periodic lensing
    signal. Since it would otherwise cause problems when performing
    Fourier transforms, we still put a periodic signal centered
    on the cluster position in the matched filter and we used the \emph{off} 
    maps to correct the bias it incurs. In order to create non periodic maps with reduced correlation between their
    edges, we created simulated maps five times larger than the
    final size, and we kept only the central part.

\begin{figure}
    \centering
    \includegraphics[width=\hsize]{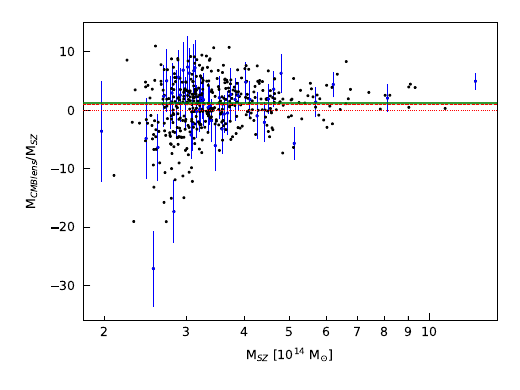}
      \caption{Recovered CMB lensing mass $M_{\rm CMBlens}$ (divided by the input fiducial mass $M_{\rm SZ}$) as a function of the input fiducial mass for the joint \Planck\ and SPT-SZ analysis of non periodic simulations. Black dots are individual measurements. Error bars are shown in blue but we display only one out of ten for clarity. The dotted (resp. dashed) red line is the zero (resp. unity) level. 
       The green stripe is the weighted mean
      $\langle M_{\rm CMBlensing}/M_{\rm SZ} \rangle = 1.19 \pm 0.18$.
              }
         \label{fig:simu_nonp}
    \end{figure}

   Results for the joint \Planck\ and SPT-SZ analysis are shown in Fig.~\ref{fig:simu_nonp} and last line of Table~\ref{tab:res_simu}. Individual measurements are shown as black dots and have large errors bars (S/N $\sim 0.1-1$) in blue. Averaging over the full sample provides $\langle M_{\rm CMBlensing}/M_{\rm SZ} \rangle = 1.19 \pm 0.18$ (green stripe) compatible with one as expected.
   
    In Table~\ref{tab:res_simu}, we also show the result for \Planck\ only and SPT-SZ only simulations in line one and two. Both results are compatible with one as expected.
    More importantly, the \Planck\ only error is $\sigma_{Planck} = 0.27$, while the SPT-SZ only error is $\sigma_{SPT} = 0.25$, close the the \Planck\ error. The joint analysis provides an error
    $\sigma = 0.19 \sim \sqrt{\frac{1}{1/\sigma_{Planck}^2+1/\sigma_{SPT}^2}}$ showing that there is indeed some additional information to gain in combining the two data sets. We investigated farther the complementarity of the two data sets in Sect.~\ref{sec:res_data}.

    \begin{table}
      \caption[]{Results on simulations, for a single realisation of a SPT-SZ-like catalogue.}
         \label{tab:res_simu}
     $$ 
         \begin{array}{p{0.5\linewidth}l}
            \hline
            \noalign{\smallskip}
            Data sets      &  M_{\rm CMBlensing}/M_{\rm SZ} \\
            \noalign{\smallskip}
            \hline
            \noalign{\smallskip}
            \Planck\ & 1.42 \pm 0.27     \\
            SPT-SZ           & 1.17 \pm 0.25 \\
            \Planck\ + SPT-SZ     & 1.19 \pm 0.18   \\
            \noalign{\smallskip}
            \hline
         \end{array}
     $$ 
   \end{table}

These runs on simulation demonstrate that our pipeline is able to extract the average CMB lensing mass of the cluster sample, and provides a joint error which is a factor of $\sim$$\sqrt{2}$ better than the error bar of individual experiments. We also made use of these simulations to assess the impact of modeling uncertainties on our results in Sect.~\ref{sec:uncer}.

\section{Results} \label{sec:res_data}

    We first present the cluster mass massurements from
    \Planck\ and SPT-SZ data separately in Sect.~\ref{sec:res_individual}.
    We then show the measurement from the joint analysis in Sect.~\ref{sec:res_joint} and discuss the improvement with respect to individual analyses.

    \subsection{Masses from individual data sets}
    \label{sec:res_individual}

    The pipeline was first used to measure the masses of the 468 clusters with the
    \Planck\ and SPT-SZ data sets separately in order to quantify the gain
    due to the combination of both.

    \subsubsection{CMB lensing masses from \Planck\ only}
   
    \begin{figure}
    \centering
    \includegraphics[width=\hsize]{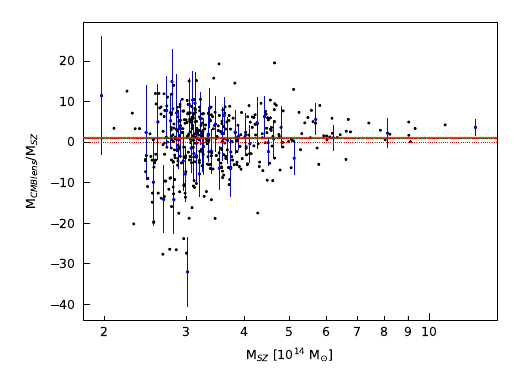}
      \caption{Recovered CMB lensing mass $M_{\rm CMBlens}$ (divided by the SZ mass $M_{\rm SZ}$) as a function of the SZ mass for the \Planck\ only analysis. Black dots are individual measurements. Error bars are shown in blue but we display only one out of ten for clarity. The dotted (resp. dashed) red line is the zero (resp. unity) level.
      The green stripe is the weighted mean
      $\langle M_{\rm CMBlensing}/M_{\rm SZ} \rangle = 1.03 \pm 0.27$.
       }
      \label{fig:planck_res}
    \end{figure}

    The result for the \Planck\ only data set is shown in Fig.~\ref{fig:planck_res}. Black dots are individual measurements, and individual error bars are shown in blue for one measurement out of ten for clarity. The weighted mean of our measurement is $M_{\rm CMBlens}/M_{\rm SZ} = 1.03 \pm 0.27$. This is a $3.7~\sigma$ measurement. It is consistent with unity, showing that the \Planck\ CMB lensing mass measurement is compatible with the SZ mass measurement $M_{\rm SZ}$, that is the SPT-SZ mass measurement multiplied by 0.8. The error bars we obtained are similar to the ones of the simulations, hinting that the \Planck\ maps are well modelled in the simulations. 
      
    \subsubsection{CMB lensing masses from SPT-SZ only}
    
      \begin{figure}
    \centering
    \includegraphics[width=\hsize]{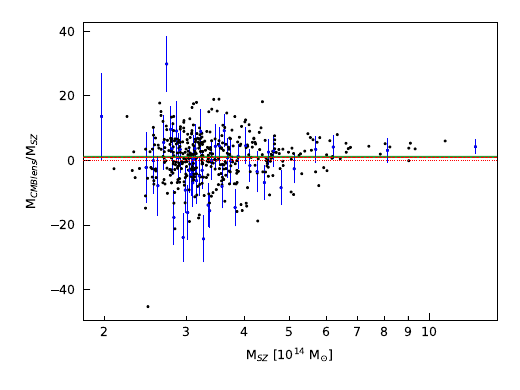}
      \caption{Recovered CMB lensing mass $M_{\rm CMBlens}$ (divided by the SZ mass $M_{\rm SZ}$) as a function of the SZ mass for the SPT-SZ only analysis. Black dots are individual measurements. Error bars are shown in blue but we display only one out of ten for clarity. The dotted (resp. dashed) red line is the zero (resp. unity) level.
      The green stripe is the weighted mean
      $\langle M_{\rm CMBlensing}/M_{\rm SZ} \rangle = 1.12 \pm	0.29$.
              }
         \label{fig:SPT_res}
    \end{figure}

    The result for the SPT-SZ only data set is shown in Fig.~\ref{fig:SPT_res}.
    We measured an average ratio $M_{\rm CMBlens}/M_{SZ} = 1.12 \pm	0.29$, a $3.9~\sigma$ measurement. It is in agreement with the \Planck\ only measurement and the SZ mass $M_{\rm SZ}$. In this case, the error bars are somewhat larger than on the simulations. This is probably due to the fact that the noise in the data is more complex than the assumption of simple white noise in the simulations.
      
    \subsubsection{Comparison of \Planck\ and SPT-SZ individual error bars}
    
\begin{figure}
    \centering
    \includegraphics[width=\hsize]{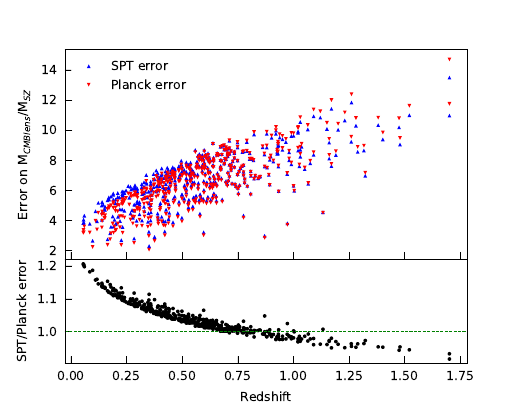}
      \caption{Comparison of the errors on the SPT and \Planck\ masses as a function of redshift $z$.
      The top panel displays the errors while
      the bottom panel displays the ratio of the errors. There is a clear
      correlation between the ratio of the errors due to the redshift dependance of the lensing potential $\Phi$.}
         \label{fig:both_err_vs_z}
    \end{figure}

    We now compare individual error bars from the \Planck\ only and the SPT-SZ only measurements. Fig.~\ref{fig:both_err_vs_z} (top panel) shows the SPT-SZ and the \Planck\ error bars as a function of redshift. The bottom panel shows the ratio of the two. \Planck\ provides error bars smaller than SPT-SZ at $z<1$, while SPT-SZ is more efficient than \Planck\ at $z>1$. However the ratio between the two remains close to unity with a factor around 1.2 at maximum (reached for the lowest redshift clusters). We note the strong correlation between the two errors bars as a function of redshift. Error bars are given by the expression $\left[ \sum_{\bm{K}} \frac{|\Phi(\bm{K})|^2}{A(\bm{K})} \right]^{-1/2}$ with $A(\bm{K})$, the variance of the Hu \& Okamoto estimator, being independent of the cluster properties (mass or redshift). Thus, at fixed mass, the redshift dependence is given by the redshift dependence of the potential $\Phi$ (or equivalently $\kappa$). So the redshift dependence of the ratio of the two errors is driven by the angular diameter distance $D_d$ in $\Sigma_{NFW}(D_d \bm{\theta})$.
     
\begin{figure}
    \centering
    \includegraphics[width=\hsize]{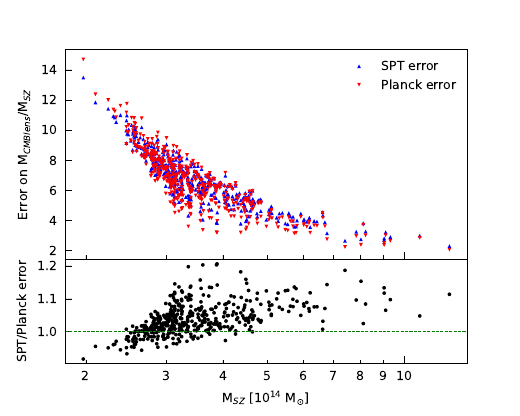}
      \caption{Comparison of the errors on the SPT and \Planck\ masses as a function of the SZ mass $M_{\rm SZ}$. The top panel displays the errors while the bottom panel displays the ratio.}
         \label{fig:both_err_vs_M}
    \end{figure}

 Fig.~\ref{fig:both_err_vs_M} shows the SPT-SZ and the \Planck\ error bars as a function of mass, and the corresponding ratio. \Planck\ outperforms SPT-SZ at high mass ($M_{\rm SZ}>3 \times 10^{14} M_\odot$) while SPT-SZ is more efficient at the low mass end.
 The dispersion of error bar ratios with respect to mass exceeds that observed for redshift dependence. This dispersion is driven by the redshift distribution of clusters at a given mass.
  
\subsection{Joint extraction of cluster masses}
\label{sec:res_joint}

\begin{figure*}
    \centering
    \includegraphics[width=\hsize]{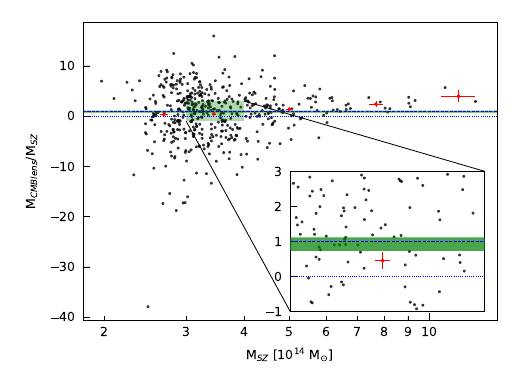}
      \caption{Recovered CMB lensing mass $M_{\rm CMBlens}$ (divided by the SZ mass $M_{\rm SZ}$) as a function of SZ mass for the \Planck\ + SPT-SZ combined analysis. Black dots are individual measurements. The red dots are the weighted means of the five mass bins, spaced logarithmically. The dotted (resp. dashed) blue line is the zero (resp. unity) level.
      The green stripe is the weighted mean within the error
      $\langle M_{\rm CMBlens}/M_{\rm SZ} \rangle = 0.919 \pm 0.190$ (without correcting for the kSZ effect).}
         \label{fig:combi_res}
    \end{figure*}

\begin{figure*}
    \centering
    \includegraphics[width=0.49\hsize]{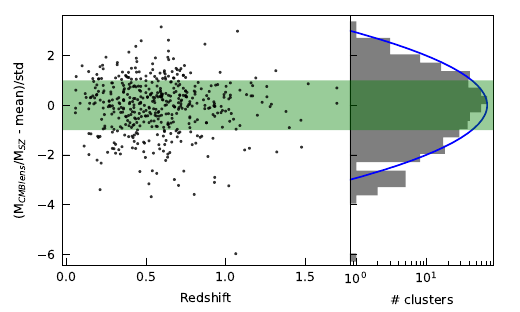}
    \includegraphics[width=0.49\hsize]{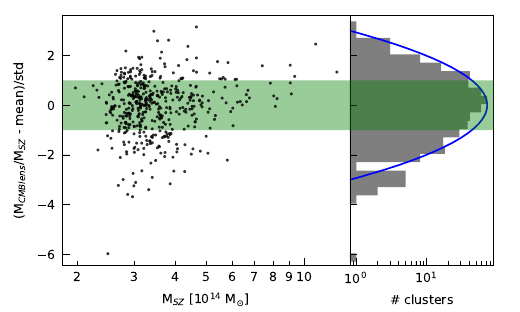}
      \caption{Pull of the ratio M$_{\rm CMBlens}$/M$_{\rm SZ}$ and associated histogram as a function of redshift ({\it Left}) and mass ({\it Right}). The blue line in each panel is a Gaussian centered on 0 with $\sigma=1$ for comparison.
              }
         \label{fig:combi_pullzm}
    \end{figure*}

    For the \Planck\ and SPT-SZ combined analysis on the 468 clusters,
    we obtained a global average ratio $M_{\rm CMBlens}/M_{\rm SZ} = 0.92 \pm	0.19$, that is a $4.8~\sigma$
    measurement of the lensing signal. As expected, the joint measurement is also in agreement with the SZ mass $M_{\rm SZ}$.
    The individual measurements are shown in Fig.~\ref{fig:combi_res}. We averaged them in five equally logarithmically spaced mass bins (red dots and associated error bars). The global average ratio is shown as a green stripe.
    The red dots present a small increasing trend with mass. We show the pull
    of the measurements in Fig.~\ref{fig:combi_pullzm}, as a function of redshift (left) and mass (right). The pull is close to a normal distribution.
    However, in the right panel, one can see the same mass trend as in Fig.~\ref{fig:combi_res}. This trend may be a hint that the $M_{\rm SZ}$ masses are overestimated at low mass and underestimated at high mass, or that there is a residual systematic bias in our measurement as a function of mass. The pull as a function of redshift (left panel) does not show any trend.

\begin{figure}
    \centering
    \includegraphics[width=\hsize]{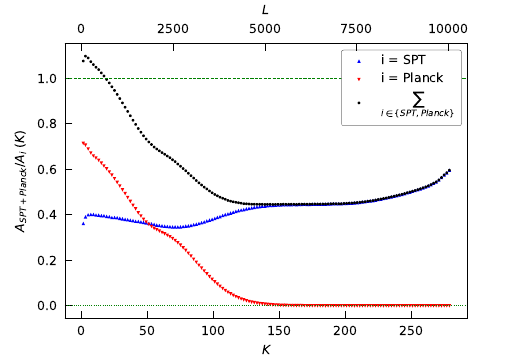}
      \caption{Comparison of the variance $A(K)$ of the lensing
      profile reconstructed for SPT-SZ and \Planck\, binned in
      $K=|\bm{k'}-\bm{k}|$ (or equivalently $L=|\bm{l'}-\bm{l}|$). The figure displays the ratios between the combined
      $A_{\rm SPT+Planck}(K)$ and the single datasets $A_i(K)$, and the sum of both ratios.
      It shows that the \Planck\ data set brings most of the information at
      large scales ($K<50.9$ or $L<1830$), while the SPT-SZ dataset provides information
      on the smaller scales ($K>145.0$ or $L>5208$). The sum is the product between
      the combined $A_{\rm SPT+Planck}$ term and the inverse variance sum $1/A_{\rm SPT}+1/A_{\rm Planck}$. It is below unity for $K>18.9$ or $L>680$ demonstrating the significant gain of the combination with respect to individual measurements.}
         \label{fig:both_A_vs_K}
    \end{figure}

    The global improvement from individual (3.7 and 3.9$\sigma$) to the joint (4.8$\sigma$) measurement is clear. In Fig.~\ref{fig:both_A_vs_K}, we aim at comparing the errors from \Planck\ only, SPT-SZ only and \Planck+SPT-SZ as a function of the spatial scale $K$ (or equivalently $L$). $A_i(K)$ is the variance of the error on the estimated lensing potential $\hat{\phi}(K)$ for the considered experiment $i$ ($i$ standing for SPT or \Planck). The ratio $A_{\rm SPT+Planck}(K)/A_i(K)$ thus compares the variance of the error of the joint measurement to the variance of the error from an individual data set measurement (SPT or \Planck). This ratio is lower than unity for SPT and \Planck\ (blue and red triangles), showing that the joint analysis provides a better measurement than individual analyses at all scales $K$. The \Planck\ dataset brings most of the information at large scales ($K<50.9$ or $L<1830$), as expected, while the SPT-SZ dataset provides information at small scales ($K>145.0$ or $L>5208$). The ranges of scales observed by SPT-SZ and \Planck\ overlap but are not identical. We also computed the sum $A_{\rm SPT+Planck}(K)/A_{\rm SPT}(K)+A_{\rm SPT+Planck}(K)/A_{\rm Planck}(K)$ (black points). It is expected to be equal to unity if the two measurements of $\hat{\phi}(K)$ from SPT and \Planck\ are independent, that is the inverse variance of the combination is the sum of the inverse variance of individual measurements. This is the case for $K=18.9$ ($L=680$). For $K< 18.9$ the sum exceeds unity showing that there is redundant information in SPT-SZ and \Planck\ data. However, the combined analysis still provides a better measurement than SPT-SZ or \Planck\ only for these modes as already noticed. More interestingly, the sum reaches values significantly below unity for $K>18.9$, showing that the combination provides a better measurement than independent measurements: the combination takes advantage of correlations between scales $\bm{k}$ from SPT-SZ and scales $\bm{k'}$ from \Planck\ to measure $\hat{\phi}(K=|\bm{k'}-\bm{k}|)$. This is an important result showing that ongoing and future ground based experiments such as Simons Observatory or CMB-S4 will still benefit from combining the dataset from their large aperture telescopes with \Planck\ to improve CMB lensing cluster mass measurements.

\section{Modeling uncertainties}
\label{sec:uncer}

The errors quoted in the previous sections are only statistical. We investigated the impact of modeling uncertainties on our result in this Sect.~\ref{sec:uncer}. We studied the impact of the kSZ in Sect.~\ref{sec:kSZ}. We considered the impact of the assumed cluster matter profile in the dedicated Sect.~\ref{sec:impact_profile}. We studied the effect of miscentering in Sect.~\ref{sec:miscen}. In Sect.~\ref{sec:red_err} and~\ref{sec:mass_err}, we studied the impact of the errors on cluster redshift and mass. Finally, we quantified the impact of the relativistic SZ effect in Sect.~\ref{sec:relat_sz}. Results are summarised in Table~\ref{tab:syst_effects} and discussed in Sect.~\ref{sec:sum_uncer}.

\subsection{Impact of the kinetic SZ}
\label{sec:kSZ}

    The spectral energy distribution of the kSZ is similar to that of the CMB. Thus, we cannot separate those two components in the ILC and we want to check the impact the kSZ has on the final result.

    We compare the results on simulated maps without kSZ, and with the kSZ signal induced by a normal velocity distribution of the galaxy clusters of 300$~$km/s, identical for all masses and redshifts. For each cluster, we performed 200 different draws of CMB, instrumental noise and velocity. We then had $200 \times 468$ measurements that we compared with $200 \times 468$ measurements with the same CMB and instrumental noise but a null velocity. On average over these 200 simulations, our result $M_{\rm CMBlens}/M_{\rm SZ}$ was shifted by $-0.060 \pm 0.015$.
    The shift is the same for different CMB realizations, with a standard deviation (0.015) four times smaller than the shift (0.060). We decided to correct our final result accordingly. We note that \cite{melin_measuring_2015} did not find a significant shift in their result due to the kSZ. This is because the shift is much smaller than the statistical error of their final result and they did not run enough simulations to extract the kSZ effect from the fluctuations of the statistical noise.

\subsection{Impact of the assumed lensing profile in the matched filter}
\label{sec:impact_profile}

    Our baseline analysis was performed with a matched filter, assuming a lensing profile based on a NFW profile truncated at $5\times r_{500}$. The real lensing profile of individual clusters scatters around the NFW profile, but one also needs to consider that clusters are not isolated objects. They are located at the nodes of the cosmic web and are thus connected to filaments and sheets. Random structures may also be present on the line-of-sight. In order to assess the impact of our profile assumption on our result, we changed the truncation radius to a lower ($3 \times r_{500}$) and a higher ($7 \times r_{500}$) value. Our result $M_{\rm CMBlens}/M_{\rm SZ}$ changed by $+0.017$ and $-0.019$ respectively (Table~\ref{tab:lens_profile}). The assumed profile thus had an impact on the result corresponding to about 10\% of the statistical error bar. Separating the effects of filaments, sheets and random structures along the line-of-sight would require the use of N-body simulations, a work which is beyond the scope of this article.
    
    \begin{table}
      \caption[]{Results on real maps changing the truncation radius used in the matched filter.}
         \label{tab:lens_profile}
     $$ 
         \begin{array}{p{0.5\linewidth}l}
            \hline
            \noalign{\smallskip}
            Truncation radius      &  \frac{M_{\rm CMBlens}}{M_{\rm SZ}} \\
            \noalign{\smallskip}
            \hline
            \noalign{\smallskip}
            $3 \times r_{500}$     & 0.936 \pm 0.202     \\
            $5 \times r_{500}$ (baseline)    & 0.919 \pm 0.190     \\
            $7 \times r_{500}$     & 0.900 \pm 0.186     \\
            \noalign{\smallskip}
            \hline
         \end{array}
     $$ 
   \end{table}

\subsection{Impact of the cluster miscentering}
\label{sec:miscen}

    In our analysis, we used the cluster positions given by~\citet{bocquet_cluster_2019}. These positions are affected by an uncertainty. For this test, we used the simulated maps for which we know the exact positions of the clusters. We drew at random 468 "observed" cluster positions from a normal law centered on the 468 "real" positions, with errors provided by Eq.~16 of~\cite{melin_measurement_2023}. We then performed the analysis on the maps, extracting the signal at the "real" and "observed" positions. The results are shown in Table~\ref{tab:misspos}. Without (resp. with) miscentering corresponds to signal extracted at "real" (resp. "observed") positions.  We measured a bias $\Delta \frac{M_{\rm CMBlens}}{M_{\rm SZ}} = +0.005$ much smaller than the error on the result $\sigma = 0.181$. We would need too many simulations to determine if the miscentering is only an increase in dispersion or if it also adds a systematic shift in the result.
    
    An alternative method allowed us to obtain the same result without simulations: we used directly the lensing profile modeled for the matched filter and we computed the matched filter between a modeled profile at the "real" and at the "observed" positions. The error used in the matched filter computation is the one we get from the estimation of the lensing profile on real maps. We did this calculation 100 times (corresponding to 100 "observed" positions) for each of the 468 clusters. With this method, we find a shift of $\Delta \frac{M_{CMBlens}}{M_{SZ}} = -0.008$, close to the one obtained on simulated maps but much more precise. There is also a dispersion due to the velocity dispersion of the clusters of $\Delta \sigma = 0.0005$ (calculated as the standard deviation across the 100 "observed" positions), negligible with respect to the shift. With this, we show that miscentering creates a small negative bias in the result. We also used this alternative method to quantify the impact of the redshift and mass errors in the next sections.
     
    \begin{table}
      \caption[]{Results on simulated maps with and without miscentering.}
         \label{tab:misspos}
     $$ 
         \begin{array}{p{0.5\linewidth}l}
            \hline
            \noalign{\smallskip}
            Miscentering     &  \frac{M_{\rm CMBlens}}{M_{\rm SZ}} \\
            \noalign{\smallskip}
            \hline
            \noalign{\smallskip}
            Without   & 1.192 \pm 0.181     \\
            With      & 1.197 \pm 0.181     \\
            \noalign{\smallskip}
            \hline
         \end{array}
     $$ 
   \end{table}

\subsection{Impact of the redshift errors}
\label{sec:red_err}

    Clusters are provided with their redshifts and associated errors. Once again, we performed the analysis on simulated maps for which we know the "real" redshifts. We drew "observed" redshifts with a normal law centered on the "real" ones with standard deviation given by the redshift uncertainty in the catalogue. The beginning of our analysis is not affected by a change in redshift because no redshift information is used until the modeling of the lensing profile in the matched filter. Unfortunately, the change in the result due to the redshift uncertainty is too small to be measured with this method. We thus used the alternative method: we used the matched filter on a lensing profile modelled with the 100 "observed" 468-redshift sets, compared to one with the "real" redshifts. The impact of the redshift errors on the matched filter is a shift of $\Delta \frac{M_{CMBlens}}{M_{SZ}} = +0.0002$ and a dispersion of about $\Delta \sigma = 0.0008$, which are negligible with respect to our statistical error $\sigma = 0.190$.

\subsection{Impact of the error on the mass $M_{\rm SZ}$}
\label{sec:mass_err}

    The direct method using simulations did not provide sufficient precision, as for the redshift. Once again, we used the alternative method, and applied the matched filter to compare the lensing profiles of the 468 clusters with "real" and "observed" M$_{500}$. We used 40 draws. We obtained a mean bias of $\Delta \frac{M_{CMBlens}}{M_{SZ}} = -0.0022$ and a dispersion of $\Delta \sigma = 0.0070$. Both are negligible with respect to the measured statistical error $\sigma = 0.190$.

\subsection{Impact of the relativistic SZ effect}
\label{sec:relat_sz}

    Our baseline result includes the relativistic SZ effect, that is $j_{\nu_i}$ in Eq.~\ref{eq:alpha_beta} is computed including relativistic corrections to the SZ effect. Details of the implementation are provided in Sect.~\ref{subsec:SZ_effect}. We removed the relativistic correction from the calculation of $j_{\nu_i}$ and re-ran the full analysis. We obtained a shift $\Delta \frac{M_{CMBlens}}{M_{SZ}} = -0.012$, of order of 10\% of the statistical error bar. In Table \ref{tab:relat_SZ}, we compare our result on real data with and without including the relativistic SZ effect.

    \begin{table}
      \caption[]{Results on real maps with and without the relativistic SZ effect.}
         \label{tab:relat_SZ}
     $$ 
         \begin{array}{p{0.5\linewidth}l}
            \hline
            \noalign{\smallskip}
            Relativistic SZ effect      &  \frac{M_{CMBlens}}{M_{SZ}} \\
            \noalign{\smallskip}
            \hline
            \noalign{\smallskip}
            With   & 0.919 \pm 0.190     \\
            Without      & 0.907 \pm 0.187     \\
            \noalign{\smallskip}
            \hline
         \end{array}
     $$ 
   \end{table}
   
\subsection{Summary of modeling uncertainties}
\label{sec:sum_uncer}

Our modeling uncertainties are summarised in Table~\ref{tab:syst_effects}. The total error (errors added in quadrature) is 0.026 corresponding to about 14\% of the statistical error ($0.190$). This demonstrates that the analyses based on the current data sets are dominated by statistical error driven by the instrumental noise of the experiments.
Note that we did not include the relativistic SZ effect in our total error for the modeling uncertainties because it is a known effect which has to be included in the baseline analysis, and we only wanted to quantify its impact on our final results. This is the reason why it is included as a separate line at the bottom of the table. Similarly, we only added the dispersion due to kSZ in the total error as we correct for the kSZ bias in the final result.

    \begin{table}
      \caption[]{Summary of the impact of the considered uncertainties on the measurement ${M_{\rm CMBlens}}/{M_{\rm SZ}}$. The major uncertainty is the profile modeling (about 10\% of the statistical error). The other uncertainties are of order of 1\%. The considered uncertainties and their modelings are described in dedicated subsections in Sect.~\ref{sec:uncer}.}
         \label{tab:syst_effects}
     $$ 
         \begin{array}{p{0.5\linewidth}l}
            \hline
            \noalign{\smallskip}
            Uncertainty     &  \Delta \frac{M_{\rm CMBlens}}{M_{\rm SZ}} \\
            \noalign{\smallskip}
            \hline
            \noalign{\smallskip}
            Dispersion due to kSZ              &    \pm 0.015 \\
            Profile up to $3\times r_{500}$     & + 0.017      \\
            Profile up to $7\times r_{500}$     & - 0.019      \\
            Miscentering              & - 0.008 \\ 
            Error on z                 & \pm 0.001 \\ 
            Error on $M_{500}$           & \pm 0.007 \\
            \hline
            {\bf Total}   &   {\bf \pm 0.026} \\
            \hline
            \noalign{\smallskip}
            Bias due to kSZ   &   -0.060  \\
            No relativistic SZ     & - 0.012      \\
            \hline
         \end{array}
     $$ 
   \end{table}

\section{Summary and conclusions}
\label{sec:conclu}

We performed the first CMB-lensing galaxy cluster mass measurement using a combination of ground and space-based surveys. We used the SPT-SZ cluster catalogue and extracted its average CMB lensing to SZ mass ratio from the SPT-SZ and \Planck\ data sets.
We found that the joint extraction outperforms the extraction on single datasets. This was not guaranteed a priori because SPT-SZ and \Planck\ data observe the same primary CMB anisotropies, so the observations could have been fully correlated, and the combination unproductive. This is not the case. We detected the average CMB-lensing signal at $4.8\sigma$ in the joint analysis compared with $3.7\sigma$ for the \Planck\ only and $3.9\sigma$ for the SPT-SZ only analysis. We measured, before correcting for the kSZ, $M_{\rm CMBlens}/M_{\rm SZ} = 0.92 \pm	0.19$ for the combination, $M_{\rm CMBlens}/M_{\rm SZ} = 1.03 \pm 0.27$ for the \Planck\ only analysis, and $M_{\rm CMBlens}/M_{\rm SZ} = 1.12 \pm 0.29$ for the SPT-SZ only analysis. For the SPT-SZ only analysis, our measured mass ratio is in agreement with the mass ratio measured by~\cite{baxter_measurement_2015} on a similar cluster sample and data set. Our final combined result, after correcting for the kSZ effect, is 

$$M_{\rm CMBlens}/M_{\rm SZ} = 0.98 \pm 0.19 \rm{~(stat.)} \pm 0.03 \rm{~(syst.)}.$$

Our measurement is dominated by the statistical error. We estimated the modeling uncertainties to be of order of 14\% of the statistical error.

We computed $M_{\rm CMBlens}/M_{\rm SZ}$, with $M_{\rm SZ}= 0.8 \times M_{\rm SPT}$ where $M_{\rm SPT}$ is the mass given by \cite{bocquet_cluster_2019}. This rescales the SPT masses to the level of \Planck\ masses. This shift between \Planck\ and SPT masses is recognized as a correction of a X-ray measurement (hydrostatic and instrumental) mass bias $b$, where the SPT analysis finds the bias to be $b=0.2$.
Since we "re-biased" the measurement, if $M_{\rm SPT}$ were an unbiased estimate of the CMB lensing mass, then our expected measurement should be $M_{\rm CMBlens}/M_{\rm SZ} = M_{\rm CMBlens}/(0.8 \times M_{\rm SPT}) = 1.25$ instead of 1. Our measurement is compatible with both $M_{\rm CMBlens}/M_{\rm SZ} = 1.25$ ($b=0.2$) and $M_{\rm CMBlens}/M_{\rm SZ} = 1$ ($b=0$).

The analysis method is based on an internal linear combination of SPT-SZ and \Planck\ maps that reconstructs the best primary CMB maps around clusters and nullifies the SZ effect. An optimal quadratic estimator is then applied to each CMB map to extract the cluster lensing potential. The last step consists in applying a matched filter to the lensing potential to measure the lensing mass and compare its value with the SZ mass. The analysis pipeline has been fully tested and characterized on simulations before being applied to the SPT-SZ and \Planck\ data sets. The major technical difficulties of the analysis consist in taking properly into account the transfer function of the ground-based experiment and dealing with the masking of point sources in the two data sets.

We showed that the gain from the ground and space-based combination with respect to analyses on individual data sets have two origins:
\begin{itemize}
\item The spatial scales observed by SPT-SZ and \Planck\ do not fully overlap, so the two experiments bring different information from different primary CMB scales. \Planck\ is more efficient at observing large scales while SPT-SZ is better at small scales.
\item The cluster lens correlates scales. Our method allows to extract information from scales observed by \Planck\ and scales observed by SPT-SZ that are correlated by the lens. This adds to the efficiency of the combination. 
\end{itemize}

These results demonstrate that both small (of order of arcmin) and large (greater than 0.5~deg) scales are important for CMB-lensing cluster studies. CMB-lensing cluster mass measurements with upcoming and future ground-based large telescopes, such as Simons Observatory and CMB-S4, will therefore benefit from a combination with \Planck\ to make full use of the lensing information.

\begin{acknowledgements}
 We would like to thank Jim Bartlett, Jacques Delabrouille, Boris Bolliet, Íñigo Zubeldia and Erik Rosenberg for fruitful discussions about the results presented in this article. We are also grateful to members of the Cosmology group of the Particule Physics Department of CEA Paris-Saclay for useful discussions while progressing on this analysis. We would also like to thank Jim Rich for his helpful comments and suggestions on parts of this article. We acknowledge the use of the Legacy Archive for Microwave Background Data Analysis (LAMBDA), part of the High Energy Astrophysics Science Archive Center (HEASARC). HEASARC/LAMBDA is a service of the Astrophysics Science Division at the NASA Goddard Space Flight Center. We also acknowledge the use of the Planck Legacy Archive. We used the HEALPix software \citep{gorski_healpix_2005} available at \url{https://healpix.sourceforge.io} and the WebPlotDigitizer~\citep{Rohatgi2022}.
\end{acknowledgements}

\bibliographystyle{aa} 
\bibliography{sptplanck_cmblens} 

\end{document}